\documentstyle[12pt,epsf,epsfig,wrapfig]{article} 
\begin{document}  
\begin{center} 
{\bfseries Single Spin Effects in Collisions of Hadrons and Heavy Ions
 at High Energy}  

\vskip 5mm 
V.V. Abramov$^{\dag}$ 

\vskip 5mm 
{\small 
 {\it 
Institute for High Energy Physics, 142281 Protvino, Moscow region, Russia
 }\\ 
$\dag$ {\it 
E-mail:  Victor.Abramov@ihep.ru
}} 
\end{center}

\vskip 5mm 
\begin{abstract} 
    Experimental data on transverse single-spin asymmetry,
 hyperon polarization and vector meson alignment in h+h, h+A and 
 A+A-collisions have been analyzed.
  A new mechanism for the origin of
 single spin effects is proposed, which takes into account the
 interaction of massive constituent quarks via their chromomagnetic
 moment with an effective inhomogeneous chromomagnetic field of strings,
 produced after the initial color exchange.  Quark spin precession in the
 color field is taken into account, which can be the reason for an
 oscillation of the single spin observables as a function of Feynman $x_{F}$ and its energy dependence.  The model predictions are
 compared with the experimental data, in particular with the heavy  ion collision data. The data are consistent with a large negative anomalous chromomagnetic moment of the constituent quarks which is predicted in the instanton model.
\end{abstract}  

\vskip 8mm 
   It is assumed in the model, that each
   quark or antiquark, which is not a constituent of the observed
   hadron $C$ in the reaction
     $ A\!\uparrow  + B  \rightarrow  C + X$ 
contributes, with some probability, to the effective color field, which acts on the hadron $C$ quarks. As is shown in \cite{Migdal}, a string arises
between the receding quark and antiquark, which has a longitudinal
chromoelectric field ${\rm\bf E^{a}}$ and a circular chromomagnetic
field ${\rm\bf B^{a}}$. The field ${\rm\bf B^{a}}$ spreads around the
string like an ordinary magnetic field surrounds a conductor with a
current:
%
\begin{equation}
 B^{(2)}_{\varphi} = -2\alpha_{s}r/\rho^{3}exp(-r^{2}/\rho^{2}),  
\label{eq:Field-B}
\end{equation}
where $r$ is a distance from the string axis, $\rho = 1.25 R_{c} =
2.08$ GeV$^{-1}$, and $R_{c}$ is the confinement radius, the index (2) in
$B^{(2)}_{\varphi}$ means a color, and $\varphi$ is the azimuthal
angle.

This inhomogeneous field $\rm\bf{B^{a}}$ acts
 on a color magnetic moment $\mu = sgq_{s}/2M_{Q}$ of the quark $Q$,
 where $q_{s}=\sqrt{4\pi \alpha_{s}}$ is the color coupling constant,
 $g$ is the color gyromagnetic number, $M_{Q}$ is the constituent quark (valon) mass.   The Stern-Gerlach-like force given by 
\begin{equation}
    f_{x} =  \mu_{x}\partial B_{x}^{a}/\partial x  +
   \mu_{y}\partial B_{y}^{a}/\partial x, 
\label{eq:force-fx}
\end{equation}
can be the reason of the large single spin asymmetry (SSA) \cite{Ryskin}.

We assume a Larmore precession \cite{BMT} of the mean quark
spin ${\rm\bf \xi}$ in the color field ${\rm\bf B^{a}}$, which depends on the quark energy $E_{Q}$:
\begin{equation}
  d{\rm\bf \xi }/dt  = a[{\rm\bf \xi B^{a} }],  
\label{eq:precession}
\end{equation}
\begin{equation}
  a = q_{s}(g - 2 + 2M_{Q}/E_{Q})/2M_{Q}.
\label{eq:precession-a}
\end{equation}

 A large negative value of the anomalous quark chromomagnetic moment 
$\mu_{a} = (g-2)/2 $ is predicted in the framework of the instanton
model:   
 $\mu_{a} = -0.744 $ (Diakonov,\cite{Diakonov}).

At high quark energies $E_{Q} \gg 2M_{Q}/|g-2|$ the quark spin
precession frequency $\Omega_{s} =aB$ is almost energy independent due
to the high $|g-2|$ value and the energy-dependent term in (\ref{eq:precession-a}) can be considered as a correction and estimated experimentally. 

 The Stern-Gerlach-like force (\ref{eq:force-fx}) produces an
 additional transverse momentum:
\begin{equation}
    \delta p_{x} 
 \approx  {{g v [1- \cos(kS)]} \over {2 \rho kS (g-2 +2M_{Q}/E_{Q})}}, 
\label{eq:delta-pt}
\end{equation}
where $kS = aB/v$ is the precession angle, $dS = vdt$, $v$ is the quark velocity, and $S$ is a
quark path length in the string field. 
 We assume that 
  $kS = \omega_{A} x_{A}$ 
in the hadron $A$ fragmentation region, or 
  $kS = \omega_{B} x_{B}$
in the hadron $B$ fragmentation region, where $\omega_{A}$ and
$\omega_{B}$ are dimensionless values and the
scaling variables are defined as
$x_{A} = (x_{R} + x_{F})/2$ and 
$x_{B} = (x_{R} - x_{F})/2$. 

The analyzing power is related with the additional $p_T$ by 
eq. (\ref{eq:ANPT}), (M.Ryskin,  \cite{Ryskin}):
\begin{equation}
A_{N} \approx \delta p_{x} \partial /\partial 
{p_{T}} \ln(d^{3}\sigma /d^{3}p).
\label{eq:ANPT}
\end{equation}
The final experession for the SSA or hadron polarization in $pp$, $pA$ or $AA$ collisions is:
\begin{equation}
A_{N} = C(\sqrt{s}) V(E_{cm}) F(p_{T},A)[G(\omega_{A}y_{A}) - 
\sigma(\theta^{cm})G(\omega_{B}y_{B})];
\label{eq:AN}
\end{equation}
\begin{equation}
G(\omega \cdot y) = [1-\cos(\omega \cdot y )]/(\omega \cdot y);
\label{eq:GX}
\end{equation}   					
\begin{equation}
\sigma(\theta^{cm}) =  \xi \sin\theta^{cm} + \varepsilon  \;;\;\;\;
F(p_{T},A)=1 -\exp[-(p_{T}/p_{T}^{min})^{3}](1-\eta \ln{A}) \;;
\label{eq:SIGMA}
\end{equation}
\begin{equation}						
y_{A} = x_{A} - (E_{0}/\sqrt{s}+f_{A})[1 +\cos\theta^{cm}] 
+ a_{0}[1 -\cos\theta^{cm}];
\label{eq:YA}
\end{equation}
\begin{equation}						
  y_{B} = x_{B} - (E_{0}/\sqrt{s}+f_{B})[1 -\cos\theta^{cm}] 
  + a_{0}[1 +\cos\theta^{cm}];
\label{eq:YB}
\end{equation}
\begin{equation}						       	    
C(\sqrt{s}) = C_{0}/(1 - E_{R}/\sqrt{s}). 
 \label{eq:CX}
\end{equation}
The Heaviside step function $V(E_{cm}) \approx \pm\Theta(E_{cm} - E_{cm}^{min})$ takes into account the threshold behavior of the SSA as a function of hadron $C$ c.m.  energy \cite{Kinem}. The
eqs. (\ref{eq:AN}) - (\ref{eq:CX}) describe not only the SSA, but also the hyperon polarization in the unpolarized
hadron collisions.   The model has 8 phenomenological parameters in the case of identical particle collision ($\omega_{A}=\omega_{B}$, 
$f_{A}=f_{B}$, $\varepsilon=1$, $\xi=0$) and 12 parameters in a general case.

  Due to the quark spin
precession (\ref{eq:precession})-(\ref{eq:precession-a}) the effective value of $E_{0}$ is given by
\begin{equation}						
 E_{0} \approx 2M_{Q} [1 + {2\over {2-g}} ],
\label{eq:E0}
\end{equation}
where it is assumed that the constituent quark mass for $u$- and $d$-quarks is the same: $M_{U}= M_{D} = 0.35$ GeV. The relation (\ref{eq:E0}) 
and the estimated values of the $E_{0}=1.640 \pm 0.040$ GeV ($\pi^{+}$)
and $E_{0}=2.02 \pm 0.21$ GeV ($\pi^{-}$)
 allow to extract the quark anomalous chromomagnetic moment for $u$- and $d$-quarks: $\mu_{a}^{U} = -0.74 \pm 0.03 (stat)$ and
 $\mu_{a}^{D} = -0.53 ^{+0.10}_{-0.07} (stat)$. These values of
  $\mu_{a}$ are compatible with the  instanton
  model prediction \cite{Diakonov}.

The hyperon polarization arises due to the Stern-Gerlach-like forces,
which separate  the spin up and down quark states  by adding a transverse
momentum to the left or to the right in the scattering plane.
The eq. (\ref{eq:AN}) predicts an oscillation of $A_{N}$ or $P_{N}$ with the 
frequency $\omega_{A}$ ($\omega_{B}$) as a function of $y_A$ ($y_B$)
and its energy dependence, eq.~(\ref{eq:CX}).

The following figures show examples of the oscillation of the SSA or hadron polarization in a wide range of energies and other kinematical variables.
The curves in the figures show the fit result  using the model function 
(\ref{eq:AN}) disscused above.
\begin{figure}[!htb]
\centering
\begin{tabular}{cc}
\begin{minipage}{77mm}
\epsfig{file=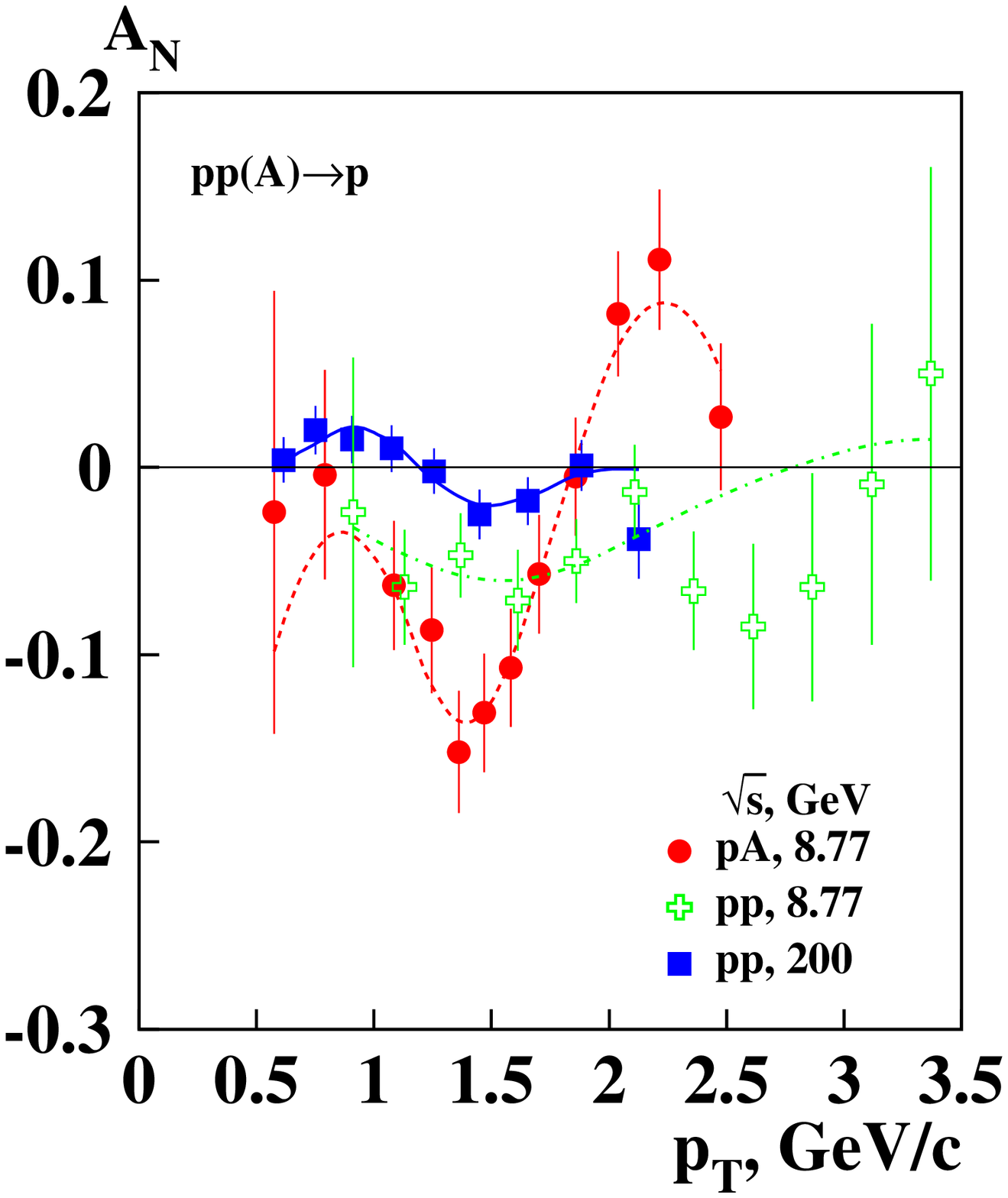,width=60mm,height=50mm}
\caption{The analyzing power vs $p_T$ for the reaction
$p\!\uparrow + p(A) \rightarrow p + X$.}
\label{PPtoP}
\end{minipage}
& 
\begin{minipage}{77mm}
\epsfig{file=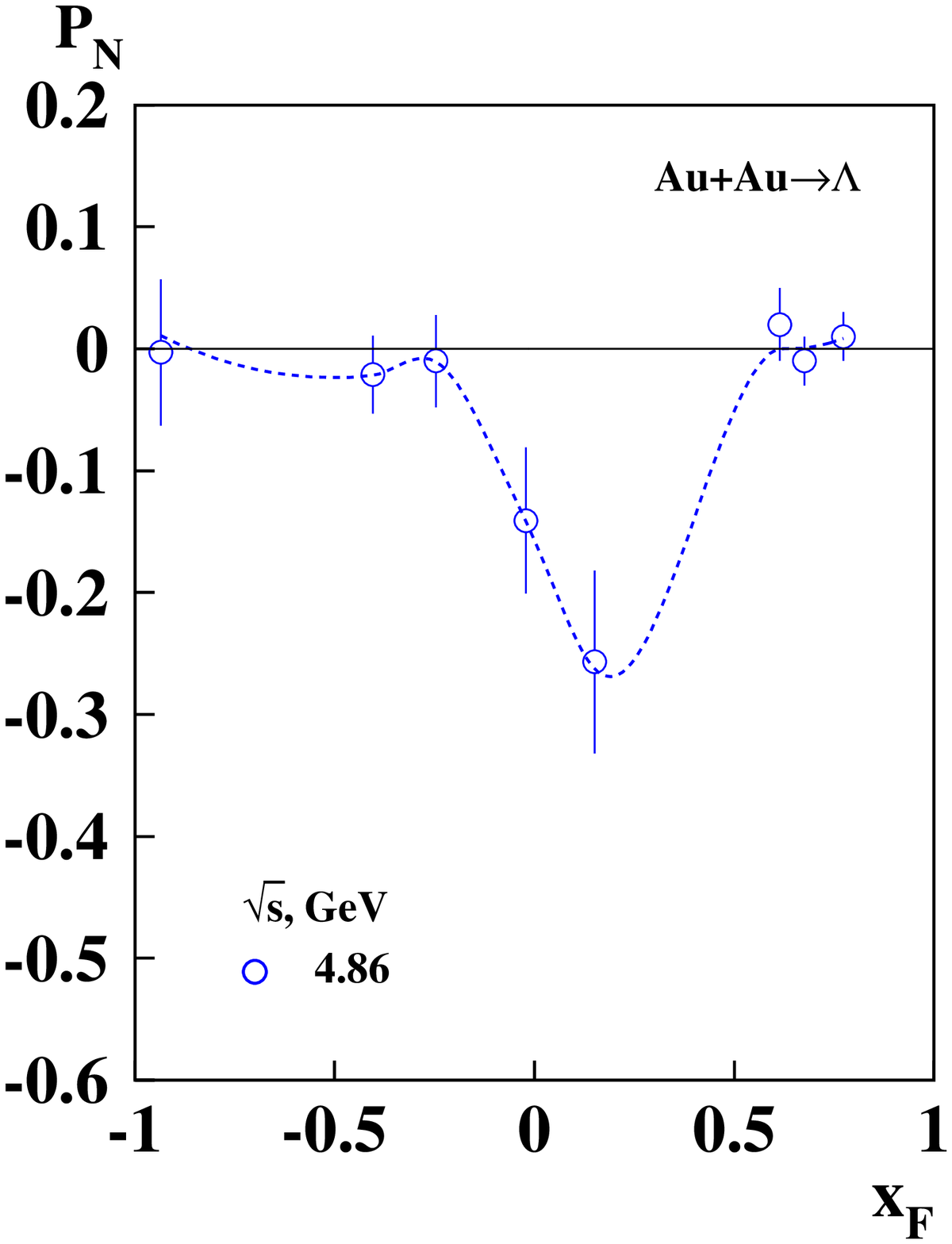,width=60mm,height=50mm}\caption{The transverse polarization vs $x_F$ for the reaction
$Au+Au  \rightarrow \Lambda\! \uparrow + X$.}
\label{AuAuE896}
\end{minipage} \\
\end{tabular}
\end{figure} 
A direct evidence of the proton $A_N$ oscillation as a function of $p_T$ 
(Fig.~\ref{PPtoP})
is obtained in the FODS-2 (IHEP) experiment using the 40 GeV/c polarized proton beam
\cite{DEG41}. Recently the $A_N$ oscillation with a smaller magnitude was observed in the BRAHMS (BNL) experiment at $\sqrt{s}=200$ GeV
\cite{BRAHMS}. The frequency $\omega_{A}$ is $-10.7 \pm 1.0$ for 
 $\sqrt{s}=8.77$ GeV and $-64 \pm 14$ for  $\sqrt{s}=200$ GeV. The rise of the $\omega_{A}$ is expected in the model due to additional sea quarks-spectators produced at high energy.

The transverse $\Lambda$ polarization in Au+Au-collisions  is measured at $\sqrt{s}=4.86$ GeV (Fig.~\ref{AuAuE896}) \cite{E896}. The fit gives positive $\omega_{A}= +18.61 \pm 0.54$, as expected in the model.
\begin{figure}[!htb]
\centering
\begin{tabular}{cc}
\begin{minipage}{77mm}
\epsfig{file=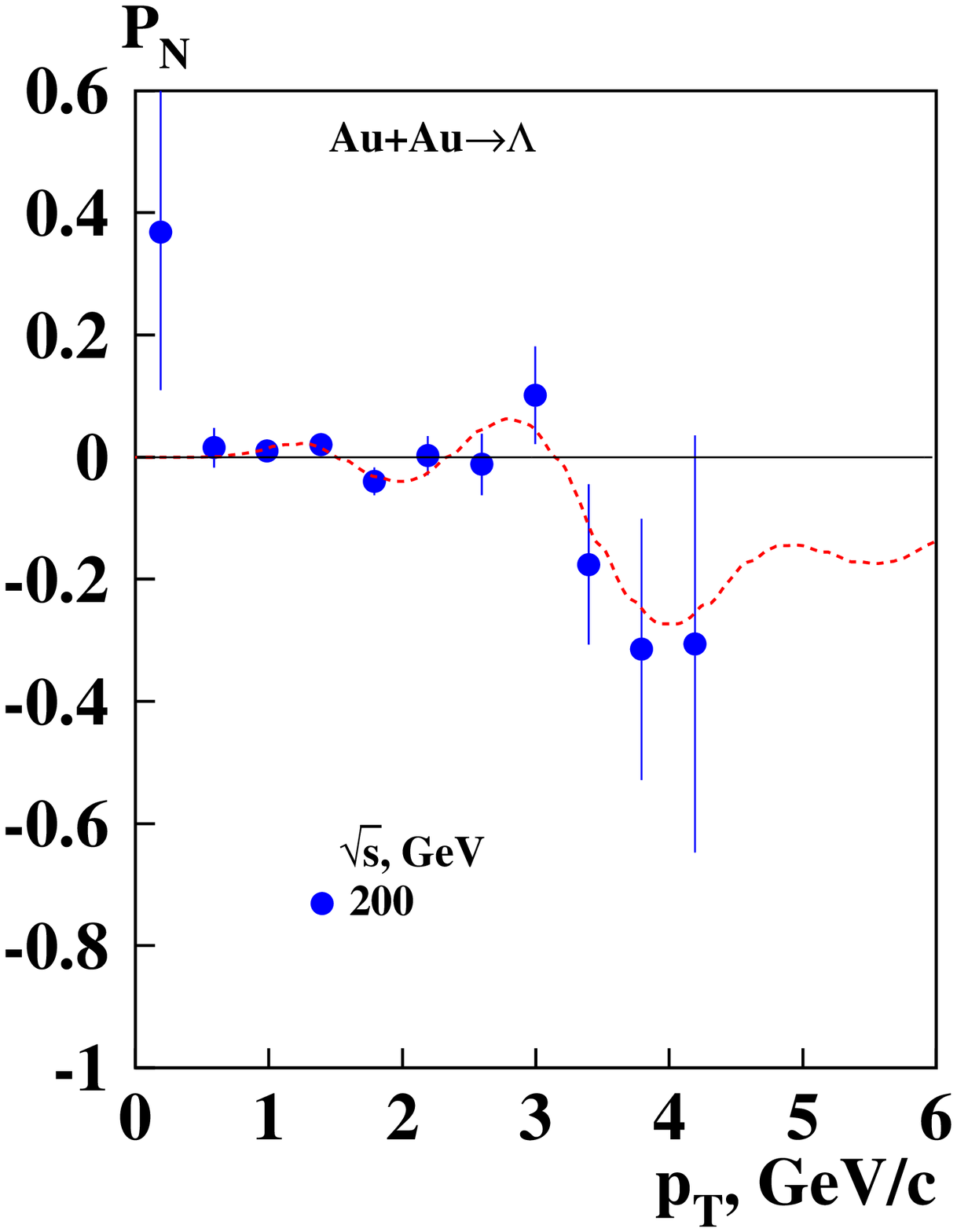,width=60mm,height=50mm}
\caption{The global $\Lambda$ polarization vs 
 $p_T$  for reaction
$Au + Au \rightarrow \Lambda\! \uparrow  + X$.}
\label{LAM200}
\end{minipage}
& 
\begin{minipage}{77mm}
\epsfig{file=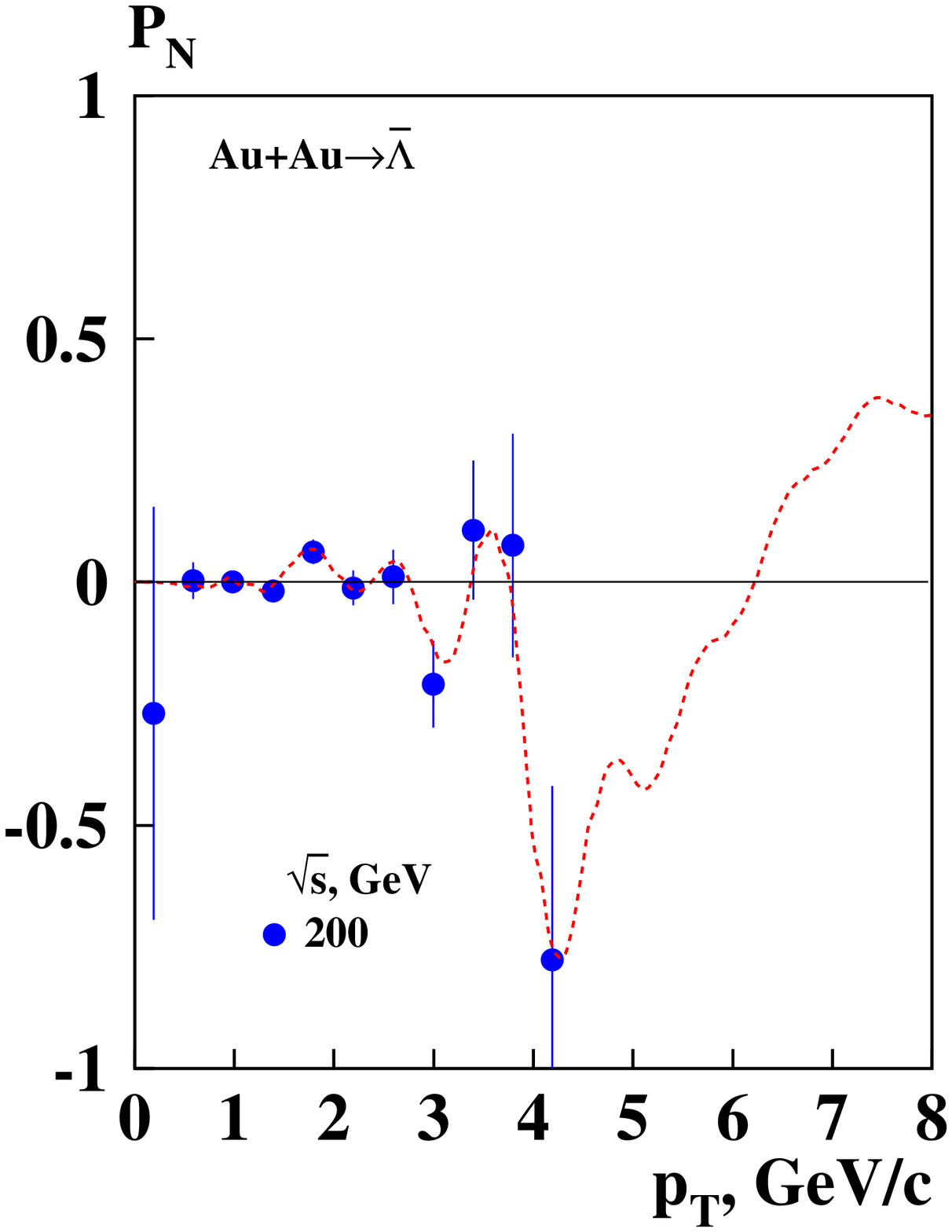,width=60mm,height=50mm}
\caption{The global $\bar{\Lambda}$ polarization vs 
 $p_T$  for reaction
$Au + Au \rightarrow \bar{\Lambda}\! \uparrow  + X$.}
\label{LAB200}
\end{minipage} \\
\end{tabular}
\end{figure} 
Recently very interesting data on the global hyperon polarization in Au+Au collisions were reported by the STAR experiment \cite{STAR}. These data show examples of polarization oscillation with negative and very high frequency $\omega_{A}$. This is exactly what is expected in the model due many spectator quarks $N_{Q} \propto A^{1/3} \exp(-w/\sqrt{s})$, whose number is proportional to the number of nucleons inside the tube of a transverse radius about the confinement radius, where $w=236 \pm 16$ GeV.
 At high reaction energy many new spectator quarks and antiquarks are produced by each nucleon that increases the field $B$. The $\Lambda$-hyperon data  fit gives  $\omega_{A} =-374 \pm 51$ for 200 GeV (Fig.~\ref{LAM200}) and $\omega_{A} =-58 \pm 38$ for 62 GeV.
The $\bar{\Lambda}$-hyperon data  fit gives  $\omega_{A} =-648 \pm 46$ and $\omega_{A} =-359 \pm 15$ for 200 GeV (Fig.~\ref{LAB200}) and 62 GeV, respectively.

\begin{wrapfigure}[16]{R}{4.8cm}
\begin{center}
\mbox{\epsfig{figure=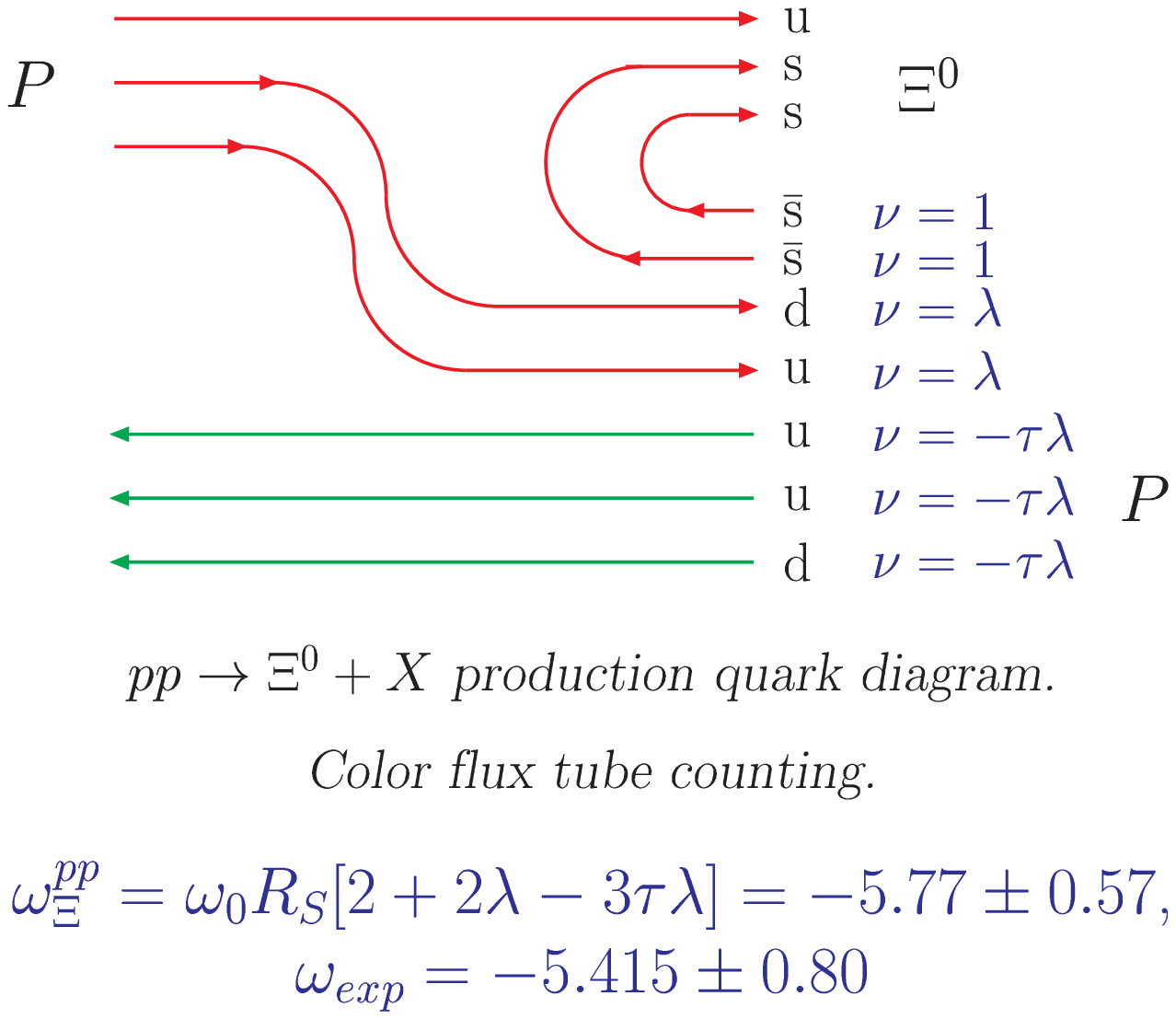,width=5cm,height=5cm}}
\end{center}
{\small{ Figure 5.} Quark counting for the reaction 
$pp  \rightarrow \Xi^{0}\! \uparrow + X$.}
\label{fxi0_pp2}
\end{wrapfigure}

The quark counting rule (QCR, Fig.~5) is designed to explain the dependence of the $\omega_{A}$ frequency on hadron quantum numbers, reaction energy and a projectile atomic weight.  The quark counting rule for the $\omega_{A}$  assumes that each projectile spectator quark contribute to the quark precession frequency with a weight $\nu=\lambda$ and each antiquark with a unity weight. For the target quarks or antiquarks an additional factor $-\tau$ should be used.
The factor $R_Q= (M_{S}/M_{Q})\mu_{a}^{Q}/\mu_{a}^{S}$ takes into account the fact that the $\omega_{A}$ is proportional to $(g-2)_{Q}/M_{Q}$. The model QCR parameters are obtained from a global fit of 26 reactions: $\omega_{0}=-3.23\pm0.30$;
$\lambda=-0.106\pm0.018$; $\tau=-0.016\pm0.027$; 
$R_{U}=1.60\pm0.24$; $R_{D}=1.95\pm0.41$; $R_{S}=1$; $R_{C}=0.78\pm0.29$.

%
%
%
Conclusion: A new mechanism is proposed which explains the origin of the
transverse single spin asymmetries and the hyperon polarization. The
origin of the single spin effects can be related with the
Stern-Gerlach-like forces between chromomagnetic moment of the massive
constituent quark and the effective color field created by the
quarks-spectators. 

The $A_N$ and $P_N$ oscillation due to the quark spin precession in the effective color field is predicted and confirmed for proton, $\Lambda$,
 $\bar{\Lambda}$, $J/\psi$, $K^{*-}(892)$, $\Xi^{0}$ and $\Xi^{-}$ production in the inclusive reactions.  The polarization oscillation is the main signature of the  model.

The estimated color anomalous magnetic moment is $-0.74 \pm 0.03$ and
$-0.53^{+0.10}_{-0.07}$ for $u$ and $d$ quark, respectively, in agreement with the instanton model prediction $\mu_{a}=-0.744$.

{\bf Acknowledgments:} We are grateful to R.Bellwied, J.C.Dunlop and I.Selyuzhenkov for
useful discussions and access to the E896 and STAR preliminary data.

%

%
%
\end{document}